\documentclass[twocolumn,floatfix,showpacs,superscriptaddress,prb]{revtex4}
\usepackage{graphicx}
\usepackage{amsmath}
\usepackage{amssymb}
\usepackage{bm}

\begin{document}

\title{Absence of a metallic phase in charge-neutral graphene with a random gap}
\author{J. H. Bardarson}
\affiliation{Materials Sciences Division, Lawrence Berkeley National Laboratory, Berkeley, CA 94720}
\affiliation{Department of Physics, University of California, Berkeley, CA 94720}
\affiliation{Laboratory of Atomic and Solid State Physics, Cornell University, Ithaca, NY 14853}
\author{M. V. Medvedyeva}
\affiliation{Instituut-Lorentz, Universiteit Leiden, P.O. Box 9506, 2300 RA Leiden, The Netherlands}
\author{J. Tworzyd{\l}o}
\affiliation{Institute of Theoretical Physics, Warsaw University, Ho\.{z}a 69, 00--681 Warsaw, Poland}
\author{A. R. Akhmerov}
\affiliation{Instituut-Lorentz, Universiteit Leiden, P.O. Box 9506, 2300 RA Leiden, The Netherlands}
\author{C. W. J. Beenakker}
\affiliation{Instituut-Lorentz, Universiteit Leiden, P.O. Box 9506, 2300 RA Leiden, The Netherlands}
\date{February 2010}
\begin{abstract}
It is known that fluctuations in the electrostatic potential allow for metallic conduction (nonzero conductivity in the limit of an infinite system) if the carriers form a single species of massless two-dimensional Dirac fermions. A nonzero uniform mass $\bar{M}$ opens up an excitation gap, localizing all states at the Dirac point of charge neutrality. Here we investigate numerically whether fluctuations $\delta M\gg\bar{M}\neq 0$ in the mass can have a similar effect as potential fluctuations, allowing for metallic conduction at the Dirac point. Our negative conclusion confirms earlier expectations, but does not support the recently predicted metallic phase in a random-gap model of graphene.  
\end{abstract}
\maketitle

Two-dimensional Anderson localization in the Dirac equation shows a much richer phase diagram than in the Schr\"{o}dinger equation.\cite{Lud94} The discovery of graphene\cite{Gei07} has provided a laboratory for the exploration of this phase diagram and renewed the interest in the transport properties of Dirac fermions.\cite{Eve08} One of the discoveries resulting from these recent investigations\cite{Bar07,Nom07,Sch09} was that electrostatic potential fluctuations $V(\bm{r})$ induce a logarithmic growth of the conductivity $\sigma\propto\ln L$ with increasing system size $L$. In contrast, in the Schr\"{o}dinger equation all states are localized by sufficiently strong potential fluctuations\cite{Lee85} and the conductivity decays exponentially with $L$.

Localized states appear in graphene if the carriers acquire a mass $M(\bm{r})$, for example due to the presence of a sublattice
symmetry breaking substrate\cite{Gio07,Zho07} or due to adsorption of atomic hydrogen.\cite{Eli09,Bos09} Anderson localization due to the combination of (long-range) spatial fluctuations in $M(\bm{r})$ and $V(\bm{r})$ appears in the same way as in the quantum Hall effect (QHE):\cite{Lud94,Nom08} All states are localized except on a phase boundary\cite{note1} of zero average mass $\bar{M}= 0$, where $\sigma$ takes on a scale invariant value of the order of the conductance quantum $G_{0}=4e^{2}/h$ (the factor of four accounts for the two-fold spin and valley degeneracies in graphene).

An altogether different phase diagram may result if only the mass fluctuates, at constant electrostatic potential tuned to the charge neutrality point (Dirac point, at energy $E=0$). The universality class is now different from the QHE, because of the particle-hole symmetry $\sigma_{x}H^{\ast}\sigma_{x}=-H$ of the single-valley Dirac Hamiltonian
\begin{equation}
H_{\rm Dirac}=v(p_{x}\sigma_{x}+p_{y}\sigma_{y})+v^{2}M(\bm{r})\sigma_{z}.\label{HDirac}
\end{equation}
This symmetry is broken by an additional electrostatic potential, as well as by intervalley scattering. Anderson localization in the presence of particle-hole symmetry has been studied extensively\cite{Cho97,Sen00,Cha02,Mil07,Kag08} in the context of superconductivity, where the Dirac spectrum appears from the superconducting order parameter rather than from the band structure. The (numerical) models used in those studies contain randomly distributed vortices in the order parameter, and are therefore not appropriate models for graphene.

It is the purpose of this work to identify, by numerical simulation, what is the phase diagram of the Dirac Hamiltonian with a random mass $M(\bm{r})=\bar{M}+\delta M(\bm{r})$ --- in the absence of any other source of disorder. This study was motivated by recent analytical work by Ziegler in the context of graphene,\cite{Zie09} which predicted a transition into a metallic phase upon increasing the disorder strength $\delta M$ at constant average mass $\bar{M}\neq 0$. Such a metal-insulator transition was known in the context of superconductivity,\cite{Sen00} but it was understood that this requires vortex disorder.\cite{Rea00,Boc00,Rea01} In order to resolve this controversy, we perform a numerical scaling analysis of the conductivity and find no metallic phase as we increase $\delta M$.

\begin{figure}[tb]
\centerline{\includegraphics[width=0.8\linewidth]{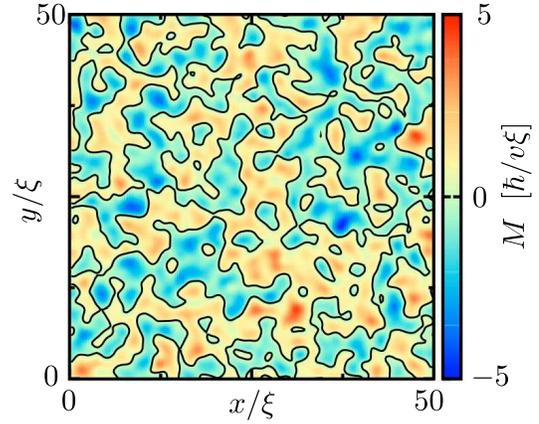}}
\caption{\label{fig_contour}
Contour plot of a random mass with Gaussian correlator \eqref{K0def}, for $K_{0}=10$. The zero-mass contours are indicated in black.
}
\end{figure}

\begin{figure}[tb]
\centerline{\includegraphics[width=0.8\linewidth]{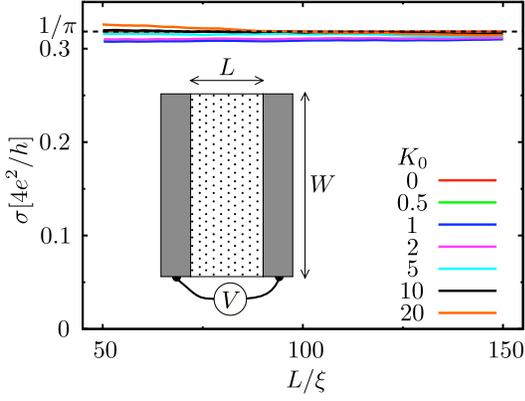}}
\caption{\label{fig_resultsa}
Average conductivity $\sigma$ as a function of length $L$ (for fixed $W=800\,\xi$). The average mass is set at $\bar{M}=0$, while the mass fluctuations are varied by varying $K_{0}$. The dashed line is at $\sigma_{0}/G_{0}=1/\pi$. The inset shows the layout of the disordered charge neutral strip (dotted rectangle) between infinitely doped electron reservoirs at a voltage difference $V$ (gray rectangles). 
}
\end{figure}

We calculate the conductivity $\sigma$ for a two-dimensional strip geometry between electron reservoirs (at $x=0$ and $x=L$, see inset in Fig.~\ref{fig_resultsa}), with periodic boundary conditions in the transverse direction (at $y=0$ and $y=W$). The Fermi level is tuned to the Dirac point in the strip, while it lies infinitely far above the Dirac point in the reservoirs. For zero mass $M$ and large aspect ratio $W/L$ the conductivity has the scale independent value\cite{Kat06,Two06} $\sigma_{0}=G_{0}/\pi$. We generate a random mass with Gaussian correlator
\begin{equation}
\langle \delta M(\bm{r})\delta M(\bm{r}')\rangle=\frac{(\hbar/v)^{2}K_{0}}{2\pi\xi^{2}}e^{-|\bm{r}-\bm{r}'|^{2}/2\xi^{2}},\label{K0def}
\end{equation}
characterized by a correlation length $\xi$ and a dimensionless strength 
\begin{equation}
K_{0}=(v/\hbar)^{2}\int d\bm{r}\,\langle \delta M(0)\delta M(\bm{r})\rangle.\label{K0def2}
\end{equation}
A contour plot for a single realization of the disorder is shown in Fig.\ \ref{fig_contour}.

\begin{figure}[tb]
\includegraphics[width=0.9\linewidth]{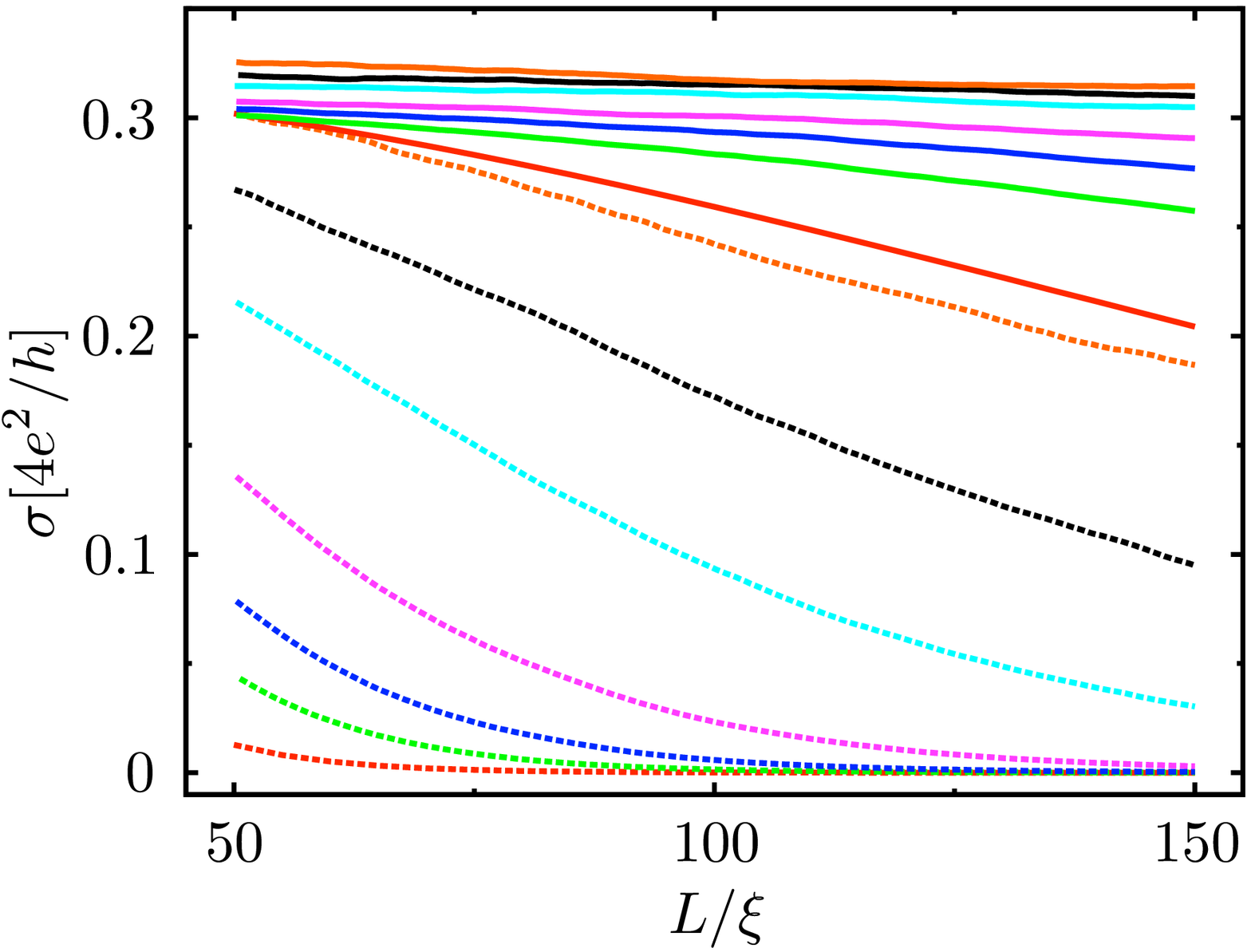}\\
\includegraphics[width=0.9\linewidth]{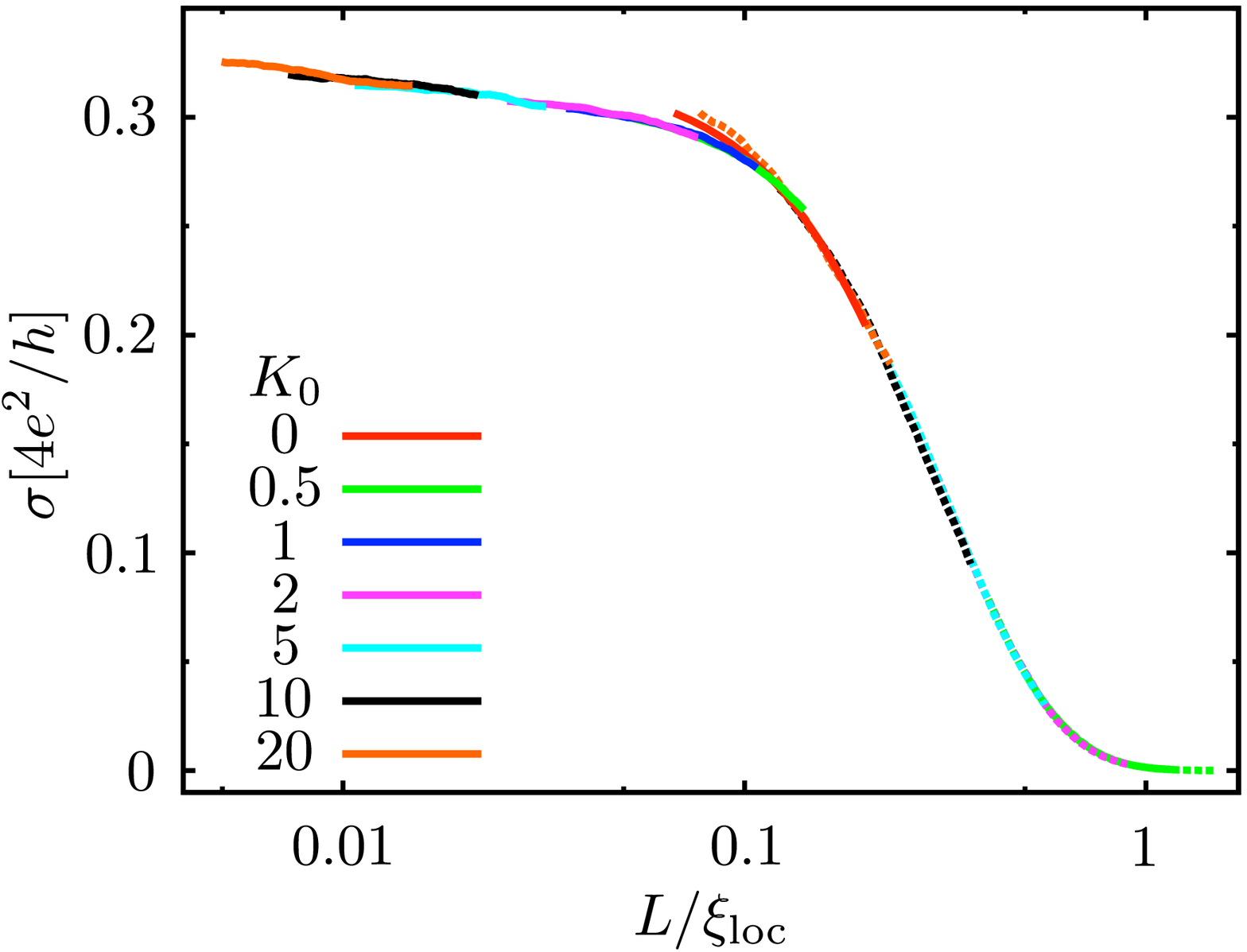}
\caption{\label{fig_resultsb}
Same as Fig.\ \ref{fig_resultsa}, but now for a nonzero average mass $\bar{M}=5\cdot 10^{-3}\,\hbar/v\xi$ (solid curves, $W=800\,\xi$) and $\bar{M}=5\cdot 10^{-2}\,\hbar/v\xi$ (dashed curves, $W=400\,\xi$). The lower panel shows the same data on a logarithmic horizontal scale, rescaled by $\xi_{\rm loc}=\xi/f(K_{0},\bar{M})$.
}
\end{figure}

The $N\times N$ transmission matrix $t$ through the strip is calculated from $H_{\rm Dirac}$ by application of the numerical method of Ref.\ \onlinecite{Bar07} to a random mass rather than to a random scalar potential. We obtain $t$ from the transfer matrix ${\cal T}$, which relates $|\psi(x=L)\rangle =
\mathcal{T}|\psi(x=0)\rangle$ and is given by
\begin{equation}
\mathcal{T} = \prod_{n=1}^{N_L}e^{\frac{1}{2}\delta xQ}\delta\mathcal{T}_ne^{\frac{1}{2}\delta xQ},\;\;
Q=-i\sigma_z\frac{\partial}{\partial y} - \frac{v}{\hbar}\bar{M}\sigma_y.
\end{equation}
Scattering from the fluctuating mass $\delta M(\bm{r})$ in the slice $(n-1)\delta x < x < n \delta x$, of incremental length $\delta x = L/N_L$, is approximated by the transfer matrix
\begin{subequations}
\label{Tapp}
\begin{align}
&  \delta\mathcal{T}_n = \frac{1-\frac{1}{2}\delta M_n(y)\sigma_y}{1+\frac{1}{2}\delta M_n(y)\sigma_y},\\
&    \delta M_n(y) = \frac{v}{\hbar}
    \int_{(n-1)\delta x}^{n\delta x} dx\, \delta M(\bm{r}).
\end{align}
\end{subequations}
The approximation \eqref{Tapp} becomes exact in the limit $N_L \rightarrow \infty$. Moreover, for any $N_L$ it satisfies the requirements of particle-hole symmetry ($\sigma_{x}{\cal T}^{\ast}\sigma_{x}={\cal T}$) as well as current conservation ($\sigma_{x}{\cal T}^{\dagger}\sigma_{x}={\cal T}^{-1}$).

We thus obtain the conductance $G=G_{0}{\rm Tr}\,tt^{\dagger}$ and the conductivity $\sigma=G\times L/W$. The number of transverse modes $N$ and longitudinal slices $N_L$ are truncated at a finite value, which is increased until a sample specific convergence is reached. For the data presented, this is typically achieved when $N = 400$---$800$ and $N_L = 300$---$600$, the larger values needed for larger values of $K_0$. The sample width $W=400\xi$---$800\xi$ is chosen large enough that the conductivity is independent of the ratio $W/L$. (Typically, $W/L\gtrsim 3$---$5$, with the larger values needed for smaller values of $\bar{M}$.) Averages over a large number of disorder configurations (typically 1000) produce the results plotted in Figs.\ \ref{fig_resultsa} and \ref{fig_resultsb}.

For $\bar{M}=0$ (Fig.\ \ref{fig_resultsa}) the conductivity stays close to the scale invariant value $\sigma_{0}$ (dashed line), no matter how large the disorder strength, while for nonzero $\bar{M}$ (Fig.\ \ref{fig_resultsb}) the conductivity decays with increasing $L$. For sufficiently large $L/\xi$ we expect single-parameter scaling, meaning that the data for different $K_{0}$ and
$\bar{M}$ should all fall on a single curve upon rescaling $L\rightarrow f(K_{0},\bar{M})L$. (This amounts to a horizontal displacement of data sets on a logarithmic horizontal scale.) The length $\xi_{\rm loc}=\xi/f(K_{0},\bar{M})$ can then be identified with the localization length (up to a multiplicative constant). As one can see in the lower panel of Fig.~\ref{fig_resultsb}, the data sets collapse reasonably well onto a single curve upon rescaling. (The remaining deviations may well be due to finite-size effects.)

For weak disorder ($K_{0}< 1$) our results are similar to earlier work on the superconducting random mass model.\cite{Cho97} That model however shows a metal-insulator transition at values of $K_{0}=K_{c}$ of order unity\cite{Cha02,Kag08} (weakly dependent on $\bar{M}$), such that for larger disorder the conductivity increases logarithmically with system size:\cite{Eve08,Sen00}
\begin{equation}
\sigma=\sigma_{0}\ln(L/\xi),\;\;{\rm for}\;\;K_{0}>K_{c}\simeq 1.\label{sigmalog}
\end{equation}
As argued by Read, Green, and Ludwig\cite{Rea00,Rea01} and by Bocquet, Serban, and Zirnbauer,\cite{Boc00} metallic conduction in a random mass landscape requires resonant transmission through contours of zero mass (the black contours in Fig.\ \ref{fig_contour}). These contours support a bound state at zero energy, if and only if they enclose an odd number of vortices. Without vortices, the phase shift accumulated upon circulating once along a zero-mass contour equals $\pi$ --- so there can be no bound state and hence no resonant transmission. (The $\pi$ phase shift is the Berry phase of the rotating pseudospin ${\bm\sigma}$ in $H_{\rm Dirac}$, without any dynamical phase shift because the energy is zero.) Our numerical finding that there is no metallic conduction in the random mass landscape without vortex disorder is therefore consistent with these analytical considerations.

From the more recent analytical work by Ziegler \cite{Zie09} we would expect a transition into a phase with a scale invariant conductivity
\begin{equation}
\sigma_{c}=\sigma_{0}[1-(\bar{M}/M_{c})^{2}\bigr],\label{Ziepred}
\end{equation}
when $M_{c}=(\hbar/v\xi)\exp(-\pi/K_{0})$ becomes larger than $\bar{M}$ with increasing disorder strength $K_{0}$. The corresponding critical disorder strength $K_{c}=\pi/\ln|v\xi/\hbar\bar{M}|\approx 0.6$---$1.0$ for the values of $\bar{M}$ in Fig.~\ref{fig_resultsb}. The numerical findings of Fig.\ \ref{fig_resultsb}, with a decaying conductivity for $K_{0}>10K_{c}$, do not support this prediction of a nonzero $M_{c}$. Note that the numerical data of Fig.\ \ref{fig_resultsa}, with a scale invariant conductivity $\sigma_{c}=\sigma_{0}$ for $\bar{M}=0$, does agree with Eq.\ \eqref{Ziepred} --- it is the $\bar{M}>0$ data that is in disagreeement.

In conclusion, we have presented numerical calculations that demonstrate the absence of metallic conduction for the Dirac Hamiltonian \eqref{HDirac}, in a random mass landscape with nonzero average and dimensionless variance $K_{0}\gg 1$. The decay of the conductivity with system size $L$ is slower for larger disorder strengths, but no metal-insulator transition is observed. A transition into a metallic phase (with $\sigma\propto\ln L$) has been attributed to vortex disorder.\cite{Rea00,Boc00,Rea01} Our numerical results are consistent with this attribution, since our model contains no vortices and has no metallic phase even if $K_{0}\gg 1$.

We have benefited from discussions with P. W. Brouwer and A. D. Mirlin. This research was supported by the Dutch Science Foundation NWO/FOM, by an ERC Advanced Investigator Grant, by the NSF under Grant No.\ DMR 0705476, and by the DOE BES. JHB thanks the Dahlem Center at FU Berlin for hospitality during the inital phase of this project.


\begin{thebibliography}{99}
\bibitem{note1} The localized phases at the two sides of the phase boundary at $\bar{M}=0$ are distinguished by the presence or absence of chiral edge states. This is similar to the QHE, but the edge states produced by a mass in graphene do not lead to a Hall voltage because they are counterprogating in the two valleys. In the computer simulations we use periodic boundary conditions, so there are no edge states and the two sides of the phase boundary are equivalent.
\end{thebibliography}
\end{document}